\documentclass[aps,pra,superscriptaddress,floatfix,showpacs,12pt,showkeys]{revtex4}
\usepackage{bbm}
\usepackage{amsfonts}
\usepackage{amsmath}
\usepackage{amssymb}
\usepackage{graphics}
\usepackage{epsfig}
\usepackage{color}
\usepackage{times}
\usepackage{graphicx}

\begin{document}

\title{Operator entanglement of two-qubit joint unitary operations revisited: Schmidt number approach}

\author{Hui-Zhi Xia}
\affiliation{Key Laboratory of Opto-electronic Information
Acquisition and Manipulation, Ministry of Education, School of
Physics {\&} Material Science, Anhui University, Hefei 230039,
People's Republic of China.}

\author{Chao Li}
\affiliation{Key Laboratory of Opto-electronic Information
Acquisition and Manipulation, Ministry of Education, School of
Physics {\&} Material Science, Anhui University, Hefei 230039,
People's Republic of China.}

\author{Qing Yang}
\affiliation{Key Laboratory of Opto-electronic Information
Acquisition and Manipulation, Ministry of Education, School of
Physics {\&} Material Science, Anhui University, Hefei 230039,
People's Republic of China.}

\author{Ming Yang \footnote{Corresponding Author: mingyang@ahu.edu.cn}}
\affiliation{Key Laboratory of Opto-electronic
Information Acquisition and Manipulation, Ministry of Education,
School of Physics {\&} Material Science, Anhui University, Hefei
230039, People's Republic of China.}

\author{Zhuo-Liang Cao}
\affiliation{School of Electronic information Engineering, Hefei
Normal University, Hefei 230061, People's Republic of China.}

\begin{abstract}
Operator entanglement of two-qubit joint unitary operations is revisited. Schmidt number is an important attribute of a two-qubit unitary operation, and may have connection with the entanglement measure of the unitary operator. We found the entanglement measure of two-qubit unitary operators is classified by the Schmidt number of the unitary operators. The exact relation between the operator entanglement and the parameters of the unitary operator is clarified too.
\end{abstract}


\date{\today}
\maketitle

\section{Introduction}
Unitary operations have been placed in a very important position in the quantum communication and entanglement manipulating, such as, quantum cryptography\cite{cryp}, teleportation\cite{tele}, entanglement swapping\cite{swapping}, quantum states purification[4], entanglement production\cite{generation} and so on. In quantum teleportation, to transfer the unknown quantum state to the remote user, the sender must apply an joint unitary operator on the unknown state particle and one of the entangled particles. In quantum entanglement swapping, a joint unitary transformation on two particles(they are from two different entangled pairs) will let two remote particles entangled without direct interaction. In entanglement purification process, joint unitary operations and measurements can transfer the entanglement from many partially entangled pairs to few near perfect entangled pairs. In entanglement generation, the joint unitary operations and single qubit operations can let the initially product particles entangled. From the above applications we can see that, it is the nonlocal attribute of the bipartite joint unitary transformation that plays the most important role. The nonlocal attribute of a bipartite joint unitary operator has been studied from different aspects, such as entangling power\cite{Zanardi1}, operator entanglement\cite{Zanardi2,wangxg1}, and entanglement-changing power\cite{Yemy}. Entangling power is the mean entanglement(linear entropy) produced by acting with U on a given distribution of pure product states\cite{Zanardi1}. Because a quantum operator belongs to a Hilbert-Schmidt space, one can consider the entanglement of the operator itself, which is named as Operator entanglement\cite{Zanardi2}. It is a natural extension of the entanglement measures of quantum states\cite{Nielsen1,Bose,Wootters,Hill} to the level of general quantum evolutions. Up to now, several methods have been proposed to quantify the entanglement of an unitary bipartite operator, such as linear entropy\cite{Zanardi2}, von Neumann entropy\cite{wangxg1}, concurrence\cite{wangxg2} and Schmidt strength\cite{Nielsen2} etc. The relations between the entangling power and these operator entanglement measures have also been discussed recently\cite{wangxg1,Zanardi2,Balakrishnan}.

In general, the entangling power of an unitary operator is related to those operator entanglement measures in complicated or indirect ways, so does the relation between the different operator entanglement measures. Clarifying the exact relation between the entangling power and different operator entanglement measures, and the relation between different operator entanglement measures will be very helpful for us to understand the nonlocal attributes and entanglement capacity of a joint unitary operator. After getting the whole nonlocal features of joint unitary operators, we can choose the optimal unitary operator to produce the specific entangled state as we want it to be, and the quantum communication protocols(such as teleportation, entanglement swapping etc) can be realized in an optimal way by introducing the appropriate joint operations\cite{future}. For two-qubit unitary operators, the two operator entanglement measures Schmidt strength and linear entropy are shown to have a one-to-one relation between them for the Schmidt number $2$ case, but no such relation exists for the Schmidt number $4$ case\cite{Balakrishnan}. This result also shows that, the Schmidt number is a very important parameter of an unitary operator when the entangling power and operator entanglement of it is concerned. In this paper, we are going to study the operator entanglement of joint two-qubit unitary operators with different Schmidt numbers. In this paper, we use the linear entropy as the entanglement measure of joint two-qubit unitary operator\cite{Zanardi2,wangxg1}, and study the Schmidt number and the entanglement measure of any unitary operator in four-dimensional Hilbert-Schmidt space. The Schmidt number of two-qubit unitary operators has the following three possible situations: $1$, $2$ or $4$\cite{Nielsen2}. We will show that the entanglement measure of two-qubit unitary operators is classified by the Schmidt number of the unitary operator. In the light of the numerical analysis, we can get the extreme value of the operator entanglement for the two-qubit unitary operators. Further, the relation between the operator entanglement and the parameters of the unitary operator will be clarified here.

\section{Operator entanglement and Schmidt number of two-qubit joint unitary operations}
There exist local unitary operators $U_{A}$,$U_{B}$ ,$V_{A}$ ,$V_{B}$
and a two-qubit unitary operator $U_{d}$, so that arbitrary two-qubit unitary operator $U_{AB}$
can be canonically decomposed as\cite{Kraus,Zhang}:
\begin{equation}
U_{AB}=(U_{A}\otimes U_{B})\cdot U_{d}\cdot(V_{A}\otimes V_{B}),
\label{decompose}
\end{equation}
where $U_{d}=\exp[-i\vec{\sigma}_{A}^{T}d \vec{\sigma}_{B}]$, and $d$ is a
diagonal matrix. In the light of this theory, any bipartite unitary operator
can be decomposed as the form above. Moreover, the entanglement measure of a unitary
operator's must be invariant under the local
unitary transformations\cite{wangxg2}. So, the entanglement measure of
any bipartite unitary operator can be simplified into the entanglement measure
of operator $U_{d}$. In the standard computational basis, we have\cite{Rezakhani}:
\begin{equation}
U_{d}=\left(
          \begin{array}{cccc}
            e^{-i{c}_{3}}c^{-} & 0 & 0 & {-i}e^{-i{c}_{3}}s^{-} \\
            0 & e^{i{c}_{3}}c^{+} & {-i}e^{i{c}_{3}}s^{+} & 0 \\
            0 & {-i}e^{i{c}_{3}}s^{+} & e^{i{c}_{3}}c^{+} & 0 \\
            {-i}e^{{-i}{c}_{3}}s^{-} & 0 & 0 & e^{{-i}{c}_{3}}c^{-} \\
          \end{array}
        \right),
        \label{decomatrix}
\end{equation}
where $c^{\pm}=\cos({c}_{1}\pm{c}_{2})$, $s^{\pm}=\sin({c}_{1}\pm{c}_{2})$, and one can always
restrict oneself to the region $\frac{\pi}{4}\geq{c}_{1}\geq{c}_{2}\geq|{c}_{3}|$, which is the so-called Weyl
chamber\cite{Zhang}.

Any operator $U$ acting on the systems $A$ and $B$ can be written in the
operator-Schmidt decomposition\cite{Horodecki}:
\begin{equation}
U=\sum_{l}{s}_{l}{A}_{l}\otimes {B}_{l},
\label{opeschmidt}
\end{equation}
where ${s}_{l}$ are the Schmidt coefficients with the positive value
and ${A}_{l}$, ${B}_{l}$ are orthonormal operator bases for $A$ and
$B$, respectively. To calculate the operator entanglement of the unitary operator $U_{AB}$, we only need to make the Schmidt decomposition of the unitary
operator ${U}_{d}$. From the Ref.\cite{wangxg1}, entanglement measure of
a unitary operator can be expressed as:
\begin{equation}
E(U)=1-\sum_{l}\frac{s_{l}^{4}}{d_{1}^{2}d_{2}^{2}},
\label{opentangle}
\end{equation}
where ${d}_{1}$ and ${d}_{2}$ are dimensions of $A$ and $B$,
respectively. So we can get the entanglement measure for the unitary
operator ${U}_{d}$ as follows:
\begin{eqnarray}
E({U}_{d})&=&1-\frac{1}{4}\{1-\sin^{2}({c}_{1}+{c}_{2})\cos^{2}({c}_{1}+{c}_{2}) -\sin^{2}({c}_{1}-{c}_{2})\cos^{2}({c}_{1}-{c}_{2})\\ \nonumber
&&+[1+2\cos^{2}(2{c}_{3})]\sin^{2}({c}_{1}+{c}_{2})\sin^{2}({c}_{1} -{c}_{2})\\ \nonumber &&+[1+2\cos^{2}(2{c}_{3})]\cos^{2}({c}_{1}+{c}_{2})\cos^{2}({c}_{1}-{c}_{2})\}.
\label{EUd}
\end{eqnarray}

The Schmidt number\cite{Nielsen1,Nielsen2} is the number of non-zero
coefficients $s_{l}$. For the unitary operator $U_{d}$, the Schmidt coefficients $s_{l}$
are as follows:
\begin{subequations}
\begin{equation}
{s}_{1}=[\cos^{2}(c_{1}+c_{2})+\cos^{2}(c_{1}-c_{2})+2\cos(2c_{3})\cos(c_{1}+c_{2})\cos(c_{1}-c_{2})]^{1/2},
\end{equation}
\begin{equation}
{s}_{2}=[\sin^{2}(c_{1}+c_{2})+\sin^{2}(c_{1}-c_{2})+2\cos(2c_{3})\sin(c_{1}+c_{2})\sin(c_{1}-c_{2})]^{1/2},
\end{equation}
\begin{equation}
{s}_{3}=[\sin^{2}(c_{1}+c_{2})+\sin^{2}(c_{1}-c_{2})-2\cos(2c_{3})\sin(c_{1}+c_{2})\sin(c_{1}-c_{2})]^{1/2},
\end{equation}
\begin{equation}
{s}_{4}=[\cos^{2}(c_{1}+c_{2})+\cos^{2}(c_{1}-c_{2})-2\cos(2c_{3})\cos(c_{1}+c_{2})\cos(c_{1}-c_{2})]^{1/2}.
\end{equation}
\label{Schmidtcoes}
\end{subequations}
We made numerical analysis for the Schmidt number and the entanglement measure of the unitary operator, and got the relation between the Schmidt number and the entanglement measure of the unitary operator(shown in Table. \ref{Table1}).
\begin{center}

\begin{table}[tbp]
\caption{The Schmidt number versus the
entanglement measure of the unitary operator ${U}_{d}$ \label{Table1}.}
\begin{ruledtabular}
\begin{tabular}{l@{}r}
\hline
Schmidt number of ${U}_{d}$ & Operator entanglement of ${U}_{d}$\\
\hline
$Sch=1$ & $E({U}_{d})=0$ \\
$Sch=2$ &  $0<E({U}_{d})\leq\frac{1}{2}$\\
$Sch=4$ &  $0<E({U}_{d})\leq\frac{3}{4}$\\
 \hline
\end{tabular}
\end{ruledtabular}
\end{table}
\end{center}

\begin{center}
\begin{figure}[tbp]
\includegraphics[width=\textwidth]{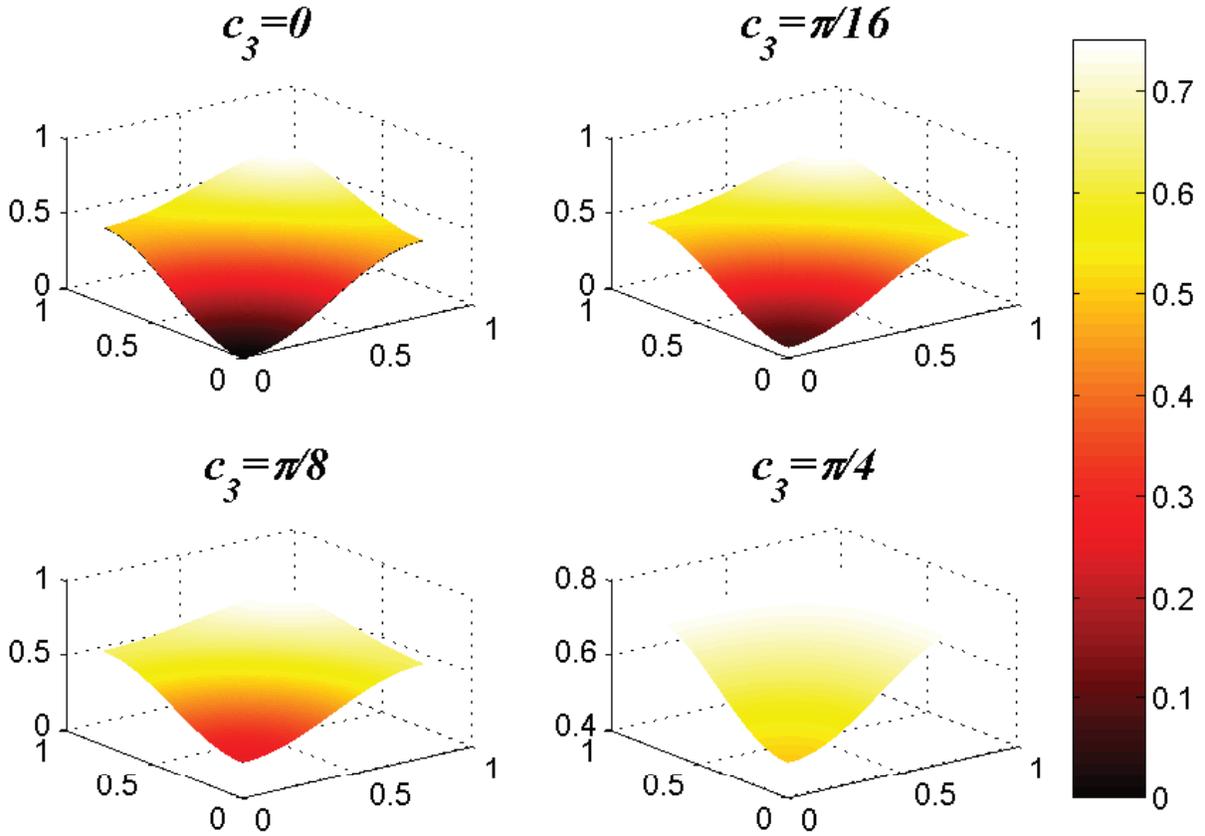}
\caption{\label{fig1}Entanglement measure of unitary operator $U_{d}$
versus the parameters $c_{1}$, $c_{2}$ for parameter
$c_{3}=0,\pi/16, \pi/8,\pi/4,$ respectively, when the Schmidt number is $4$.}
\end{figure}
\end{center}

For the Schmidt number $4$ case, the first plot in Fig.\ref{fig1} shows how the entanglement measure of $U_{d}$ depends
on the parameters $c_{1}$ and $c_{2}$ for $c_{3}=0$ when the
Schmidt number of $U_{d}$ is $4$. As the parameters $c_{1}$ and
$c_{2}$ approach to $0$, which represents a unit matrix, the
entanglement measure of the unitary operator approaches to $0$. As the
parameters $c_{1}$ and $c_{2}$ are equal to $\frac{\pi}{4}$, which
represents the SWAP gate, the entanglement measure of the unitary
operator is equal to the maximum value $\frac{3}{4}$. As the
parameters ${c}_{1}$ and ${c}_{2}$ are increasing, the entanglement
measure of unitary operator $U_{d}$ is increasing too. If $c_{3}\neq 0$, the changing pattern of the operator entanglement is similar to that of the $c_{3}=0$ case. From the other three plots in Fig.\ref{fig1}, we can see that the minimum entanglement for the Schmidt-$4$ operator is oscillating with $c_{3}$, and the maximum value and the period of the oscillation are $0.5$ and $\frac{\pi}{2}$, respectively. When $c_{3}=\frac{\pi}{4}$, the minimum entanglement reaches its maximum $0.5$. The maximum value of operator entanglement is still $\frac{3}{4}$.

\begin{center}
\begin{figure}[tbp]
\includegraphics[width=\textwidth]{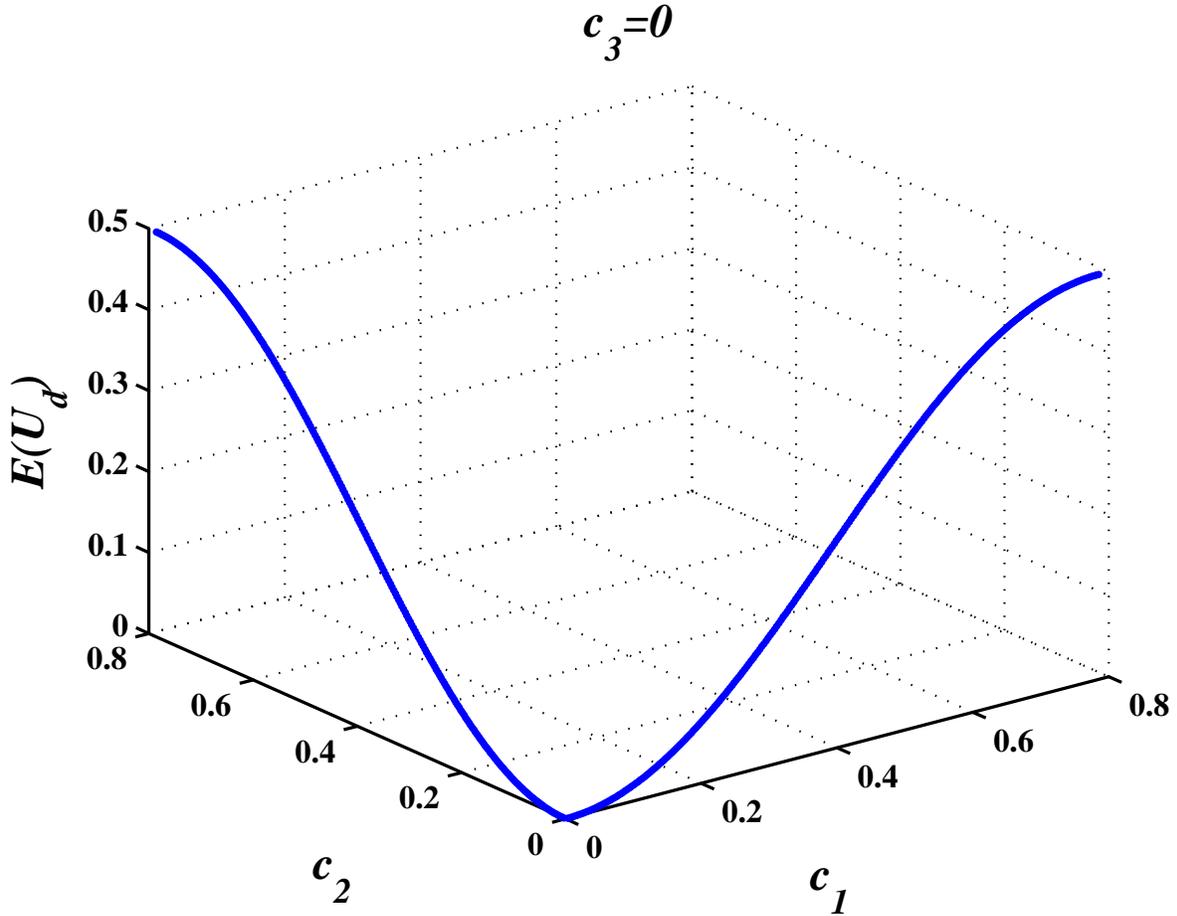}
\caption{\label{fig2}Entanglement measure of unitary operator $U_{d}$
versus the parameters $c_{1}$, $c_{2}$ for parameter
${c}_{3}=0$ when the Schmidt number is $2$.}
\end{figure}
\end{center}

For the Schmidt number $2$ case, if $c_{3}\neq0 $, then $c_{1},c_{2}$ must be zero, so the operator entanglement can be expressed in a very simple form $E({U}_{d})=\frac{1}{2}\sin^{2}(2c_{3})$. Fig.\ref{fig2} shows how the entanglement measure of $U_{d}$ depends
on the parameters $c_{1}$ and $c_{2}$ for $c_{3}=0$  when the
Schmidt number of $U_{d}$ is $2$. This curve is the boundary line of the first plot in Fig.\ref{fig1}. As the parameters $c_{1}$ and $c_{2}$  approach to $0$, the entanglement measure of unitary operator $U_{d}$ approaches to $0$. The entanglement measure of unitary operator will increase as the parameters $c_{1}$  or $c_{2}$ increase. The entanglement measure of the unitary operator with Schmidt number $2$ can get to the extreme value $\frac{1}{2}$.

From the two figures we can see that, if we want to design an operation so that it has a specific operator entanglement(or entangling power), we have infinite design schemes(i.e. $c_{1},c_{2},c_{3}$) for the Schmidt-number-$4$ type operations. But we only have two design schemes for the Schmidt-number-$2$ type operations. That is to say, Schmidt-number-$4$ type operations have a variety of design schemes rather than the only two design schemes of the Schmidt-number-$2$ type operations. So, the Schmidt-number-$4$ type operations are superior to the Schmidt-number-$2$ type operations. In addition, the maximum operator entanglement of the Schmidt-number-$4$ type operations can reach $\frac{3}{4}$, while the maximum operator entanglement of the Schmidt-number-$2$ type operations only reaches $\frac{1}{2}$. So, in practice, we will prefer the Schmidt-number-$4$ type operations.

\section{An example in Cavity QED system}

\begin{center}
\begin{figure}[tbp]
\includegraphics[width=\textwidth]{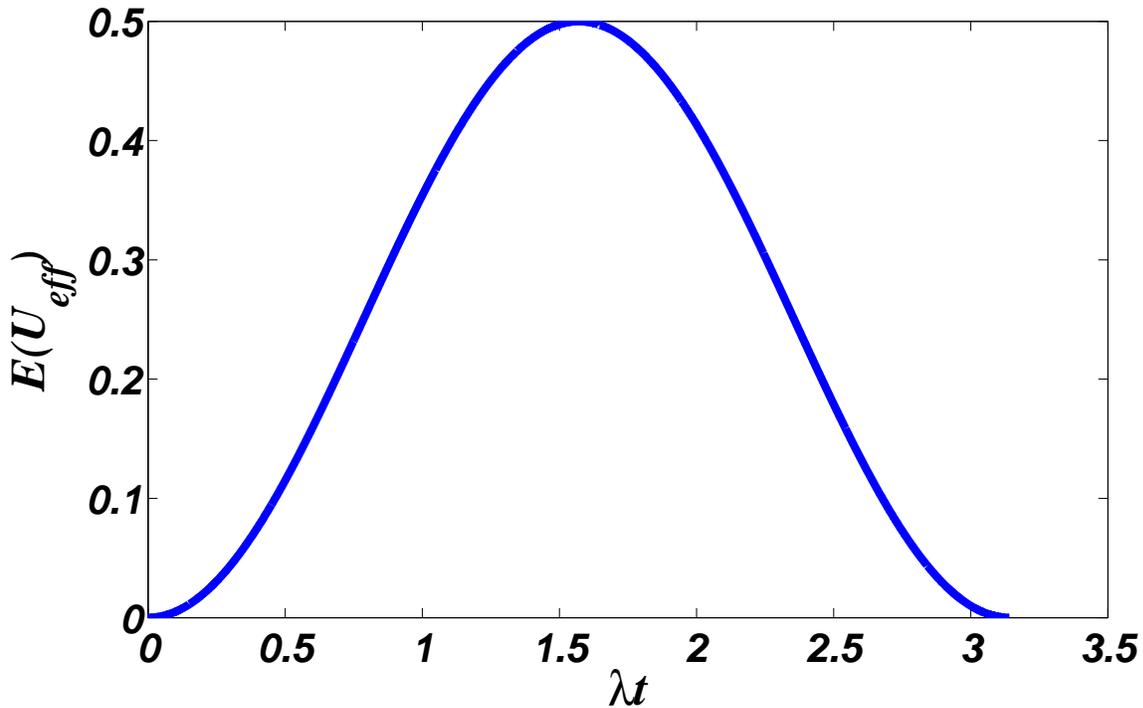}
\caption{\label{fig3}Entanglement measure of unitary operator
$U_{eff}$ versus the effective interaction time $\lambda t$ between the two atoms. Here the Schmidt number of $U_{eff}$ is $2$.}
\end{figure}
\end{center}

To demonstrate the abstract relation between the operator entanglement and the parameters of the unitary operator, we will take the following detailed example. Consider two two-level atoms($1$, $2$) trapped in a single-mode optical cavity, and the two atoms are coupled to the cavity mode with the same coupling constant $g$. The excited state$|e\rangle_{i}$ and the ground state$|g\rangle_{i}$, $(i=1,2)$ are the two levels used to encode quantum information. The two atoms have different transition frequencies $\omega_{1}$, $\omega_{2}$, and $\omega_{1}\neq \omega_{2}$. The frequency of the cavity mode is denoted by $\omega_{0}$. The atom $1$ is resonantly driven by an external classical field with coupling constant $\Omega$. Suppose the cavity mode is initially prepared in vacuum state, under the large detuning condition $\delta_{1}=\omega_{1}-\omega_{0}\gg g$, $\delta_{2}=\omega_{2}-\omega_{0}\gg g$ and in the strong driving regime $\Omega\gg \frac{g^{2}}{\delta_{1}}$, the effective Hamiltonian of the total system can be expressed as\cite{Song2010}:

\begin{equation}\label{Heff}
H_{eff}=\frac{\lambda}{2}\sigma_{1}^{x}\sigma_{2}^{x},
\end{equation}
where $\lambda=\frac{g^{2}}{\delta_{1}}$ is the effective coupling constant between atoms $1$ and $2$, and $\sigma_{i}^{x}$ is the Pauli operator of the $i$th atom. The unitary transformation induced by this effective Hamiltonian can be expressed as:
\begin{equation} U_{eff}=\left(
          \begin{array}{cccc}
            cos(\frac{\lambda t}{2}) & 0 & 0 & {-i}sin(\frac{\lambda t}{2}) \\
            0 & cos(\frac{\lambda t}{2}) & {-i}sin(\frac{\lambda t}{2}) & 0 \\
            0 & {-i}sin(\frac{\lambda t}{2}) & cos(\frac{\lambda t}{2}) & 0 \\
            {-i}sin(\frac{\lambda t}{2}) & 0 & 0 & cos(\frac{\lambda t}{2}) \\
          \end{array}
        \right).
        \label{Ueff}
\end{equation}
If we set $c_{1}=\frac{\lambda t}{2}$, ${c}_{2}=0$, ${c}_{3}=0$ in Eq.(\ref{decomatrix}), it is just the joint unitary operator in Eq.(\ref{Ueff}). That is to say, the above mentioned physical process is just a physical realization of the joint unitary operation (\ref{decomatrix}). The Schmidt number of the operator (\ref{Ueff}) is $2$, and the operator entanglement measure of it can be expressed as $E(U_{eff})=\frac{1}{2}\sin^{2}(\lambda t)$. The relationship between the operator's entanglement measure and the effective interaction time $\lambda t$ between the two atoms is depicted in Fig.\ref{fig3}. From this figure we can easily see that the maximum operator entanglement is $\frac{1}{2}$ with $Sch=2$.

\section{Conclusion}

In this paper, the linear entropy and the Schmidt number of an arbitrary two-qubit unitary operator is discussed. The results have
shown that the Schmidt number is related with the entanglement measure of unitary operators closely. For the same operator entanglement within the range $(0,\frac{1}{2}]$, there exist infinite unitary operators with Schmidt number
$4$ but only $2$ unitary operators with Schmidt number $2$. In this sense, we can say that the unitary operators with Schmidt
number $4$ can be realized more easily than the unitary operators with Schmidt number $2$ if the same operator entanglement is required. In addition, for the unitary operators with Schmidt number $4$, the range for the operator entanglement is $(0,\frac{3}{4}]$. But, for the unitary operators with Schmidt number $2$, the range decline to $(0,\frac{1}{2}]$. There must be some requirement of entanglement which can be available for the unitary operator with Schmidt number $4$ only.

\section*{Acknowledgments}

This work is supported by National Natural Science Foundation of China (NSFC) under Grants No. 10704001, No. 61073048, No.10905024 and 11005029, the Specialized Research Fund for the Doctoral Program of Higher Education(20113401110002), the Key Project of Chinese Ministry of Education.(No.210092), the China Postdoctoral Science Foundation under Grant No. 20110490825, Anhui Provincial Natural Science Foundation under Grants No. 11040606M16 and 10040606Q51, the Key Program of the Education Department of Anhui Province under Grants No. KJ2012A020, No. KJ2012A244, No. KJ2010A287, No. KJ2012B075 and No. 2010SQRL153ZD, the `211' Project of Anhui University, the Talent Foundation of Anhui University under Grant No.33190019, the personnel department of Anhui province, and Anhui Key Laboratory of Information Materials and Devices (Anhui University).


\begin{thebibliography}{99}
\bibitem[]{future}The optimality of these quanutm communication protocols by introducing the appropriate joint unitary operations will be discussed in another work which will be published soon.
\end{thebibliography}
\end{document}